\documentclass[article]{JHEP3}

\usepackage{graphicx}
\usepackage{amsmath,epsfig}
\usepackage{amssymb,amsfonts}
\usepackage{latexsym}

\relax

\def\be{\begin{equation}}
\def\ee{\end{equation}}

\newcommand{\bear}{\begin{eqnarray}}
\newcommand{\bea}{\begin{eqnarray}}
\newcommand{\eear}{\end{eqnarray}}
\newcommand{\eea}{\end{eqnarray}}
\def\hri#1#2{\href{http://arxiv.org/abs/#1}{[ArXiv:#1]#2}}
\def\hre#1#2{\href{http://arxiv.org/abs/#1/#2}{[ArXiv:#1/#2]}}

\newbox\pippobox

\def\II{\relax{\rm I\kern-.18em I}}

\def\e{\epsilon}
\def\l{\lambda}

\def\sp{\;\;\;,\;\;\;}

\title{Mini-Black-Hole production at RHIC and LHC.}

\author{\href{http://hep.physics.uoc.gr/~kiritsis/}{Elias Kiritsis}$^{a,b}$, Anastasios Taliotis$^a$\\
~\\
$^a$ \href{http://hep.physics.uoc.gr}{Crete Center for Theoretical Physics},
Department of Physics, University of Crete, 71003 Heraklion, Greece\\
~\\
$^b$ \href{http://www.apc.univ-paris7.fr}{APC, Universit\'e Paris 7}, \\ B\^atiment Condorcet, F-75205, Paris Cedex 13, France (UMR du CNRS 7164).}

\preprint{CCTP-2011-35}

\abstract{We argue that heavy-ion collisions provide the best testing ground for mini-black hole physics as $M_P\simeq 4 GeV$ for the gravity dual of YM and give concrete evidence for a new extra dimension, that is visible only to the strong interactions.
We analyse the process of production evolution and decay of the mini-black-holes by using recent results on gravity duals of YM. There are several novelties compared with the traditional story of black hole evaporation, including Bjorken scaling instead of sphericity, evaporation via bubble nucleation instead of the Hawking mechanism and lepton-poor final states. Multiplicities are estimated using shock-wave scattering techniques.
It is argued that high-multiplicity/high energy pp collisions will also show similar characteristics of mini-black-hole production and decay.
}

\keywords{ Gauge-gravity correspondence, Tachyon Condensation, QCD Anomaly, Spontaneous Symmetry Breaking}

\dedicated{Based on talks given at the Planck 2011 conference in Lisboa, Portugal and the EPS conference in Grenoble, France, July 2011. To appear in the proceedings of the EPS conference.}

\begin{document}

\maketitle 

\section{Introduction}

The subject of the EPS talk was a review of holographic model of four dimensional YMs, \cite{ihqcd}, its implications for the finite temperature
structure of the theory, \cite{gkmn}, and its predictions for the calculation of transport coefficients \cite{transport} and the Langevin diffusion of heavy quarks in the quark-gluon plasma, \cite{langevin-1}. A review of the model can be found in \cite{ihqcdrev}.

What we would like to do here,  is to explore the physics of the heavy-ion collisions in the dual gravitational description and argue that this process can be described in the dual gravitational description as the collision of two energy lumps that form a trapped surface that eventually leads to a horizon formation, and ensuing thermalization. The mini-black hole thus produced is unstable and it evolves classically until it decays via evaporation to a large multiplicity of final states.

Such a picture was explored in the context of an AdS theory with a hard IR cutoff by a number of authors, \cite{nastase1,aharony}. Since then, the structure of the bulk theory is better understood, and we will analyze the process in view of the new developments.

Black holes have dominated the mysteries associated with gravity over the past decades. It therefore seemed like a golden opportunity when a production of mini-black-holes seemed credible, \cite{dvali,giddings} at energies not far from the TeV region, if gravity had a lower characteristic scale.
This provided hope that the mysteries of black-holes, namely their thermodynamics behavior, their entropy of extreme size, the Hawking radiation and the ensuing information paradox, could be studied at colliders.
Experimenters could provide an experimental scrutiny that looks more promising  that expecting indirect signals from astrophysical black holes that are ``stars" these days of galaxy-core gravitational phenomena.

Theories of large extra dimensions are still candidates for reality, however current experimental constrains make the size of a higher-dimensional Planck scale recede further from the currently accessible energy range, making the possibility of mini-black hole production look rather remote.

Despite this unfortunate state of affairs, the gauge theory/string theory correspondence provided an alternative arena for mini-black hole production.
QCD, the gauge theory of strong interactions is expected at large $N_c$ to be described by a string theory. This theory seems elusive, as the string is very soft at high energies, but the expectation of a gravitational description at low energies is well founded, and a model that has the right properties and matches many YM observables has been proposed in \cite{ihqcd} and \cite{gubser}.

We will argue further on, that the heavy-ion collisions at CERN and LHC as well as the proton-proton collisions at LHC with high-multiplicity final states can be described by the formation and evaporation of a mini-black-hole, possible as the 5-dimensional Planck scale has been determined to be $M_{P}\simeq 4$ GeV, \cite{data}.

Moreover,  the gravitational theory has a single extra dimension, \cite{ihqcd}, that is indirectly visible via the gravitational physics. This dimension may be expected to  be a bit "fuzzy" as the the number of colors of the strong interactions $N_c=3$ seems far from the large-$N_c$ limit. However,  as lattice simulations suggest (see figure \ref{fig1} and \cite{Panero}),  SU(3) YM seems very close to SU($\infty$) YM. It is therefore fair to say that most of the quantum gravitational effects, controlled by $1/N_c^2$ are small.

The process of black hole formation and evaporation  in the dual theory of QCD has similarities with expectations for standard asymptotically flat black-hole production in gravity, like

\begin{enumerate}

\item  large total cross section,

\item very large multiplicity events, but with few hard final particles

\item suppression of hard perturbative scattering processes,

  \item Democratic primary decay (but only in the strong interactions).

  \end{enumerate}

However, the process of black hole formation and evaporation  in the dual theory of QCD  differs in several respects from the processes that many authors advocated for LHC, namely

\begin{enumerate}

\item The gravity in question is not the standard gravity, but the one of QCD, where the gravity is mediated by the $2^{++}$ glueball. Other fields, and most importantly the ``dilaton" ($0^{++}$ glueball) play also an important role.

\item  Instead of high sphericity events, there is approximate Bjorken boost invariance in the metric.

\item  They are BHs in asymptotically AdS space (almost) instead of
asymptotically flat space.

\item  Their primary decay is not via Hawking radiation but via bubble nucleation.

\item  The ratio of hadrons to leptons is small.

\end{enumerate}
The main reason for the differences is that  the BHs thus generated are live in a curved space, and
that the theory contains also other fields (in particular the dilatonic scalar).

\section{The static, translationally-invariant black holes of IHQCD}

These are the translationally invariant saddle-point  solutions, of IHQCD, defined by the Einstein-dilaton Lagrangian, \cite{ihqcdrev},
\begin{equation}
S_5=-M^3_pN_c^2\int d^5x\sqrt{g}
\left[R-{4\over 3}(\partial\phi)^2+V(\phi) \right]+2M^3_pN_c^2\int_{\partial M}d^4x \sqrt{h}~K.
 \label{kira1}\end{equation}
Here, $M_p$ is the  five-dimensional Planck scale and $N_c$ is the number of colors.
The last term is the Gibbons-Hawking term, \index{Gibbons-Hawking term} with $K$ being the extrinsic curvature
of the boundary. The effective five-dimensional Newton constant
is $G_5 = 1/(16\pi M_p^3 N_c^2)$, and it is small in the large-$N_c$ limit.

Of the 5D coordinates $\{x_i, r\}_{i=0\ldots 3}$, $x_i$ are identified with the
4D space-time coordinates, whereas  the  radial coordinate $r$ roughly corresponds to the 4D RG scale.
We identify $\l\equiv e^\phi$ with the  running 't Hooft  coupling $ N_cg_{YM}^2$,
up to an {\it a priori} unknown multiplicative factor.

The dynamics is encoded in the dilaton potential,  $V(\l)$.
The small-$\l$ and large-$\l$ asymptotics of $V(\l)$ determine the solution in the UV
and  the IR of the geometry
respectively.
\begin{enumerate}
\item For small $\l$,   $V(\l)$  is required to have a power-law expansion of the form:
\be \label{kirUVexp}
V(\l) \sim {12\over \ell^2}(1+ v_0 \l + v_1 \l^2 +\ldots), \qquad \l\to 0 \;.
\ee

\item For large $\l$,   confinement and linear Regge trajectories ($m_n^2\sim n$)  require:
\be\label{kirIRexp}
 V(\l) \sim \l^{4\over 3}\sqrt{\log \l} \quad \l\to \infty,
\ee

\end{enumerate}

\begin{figure}
\includegraphics[width=8.cm]{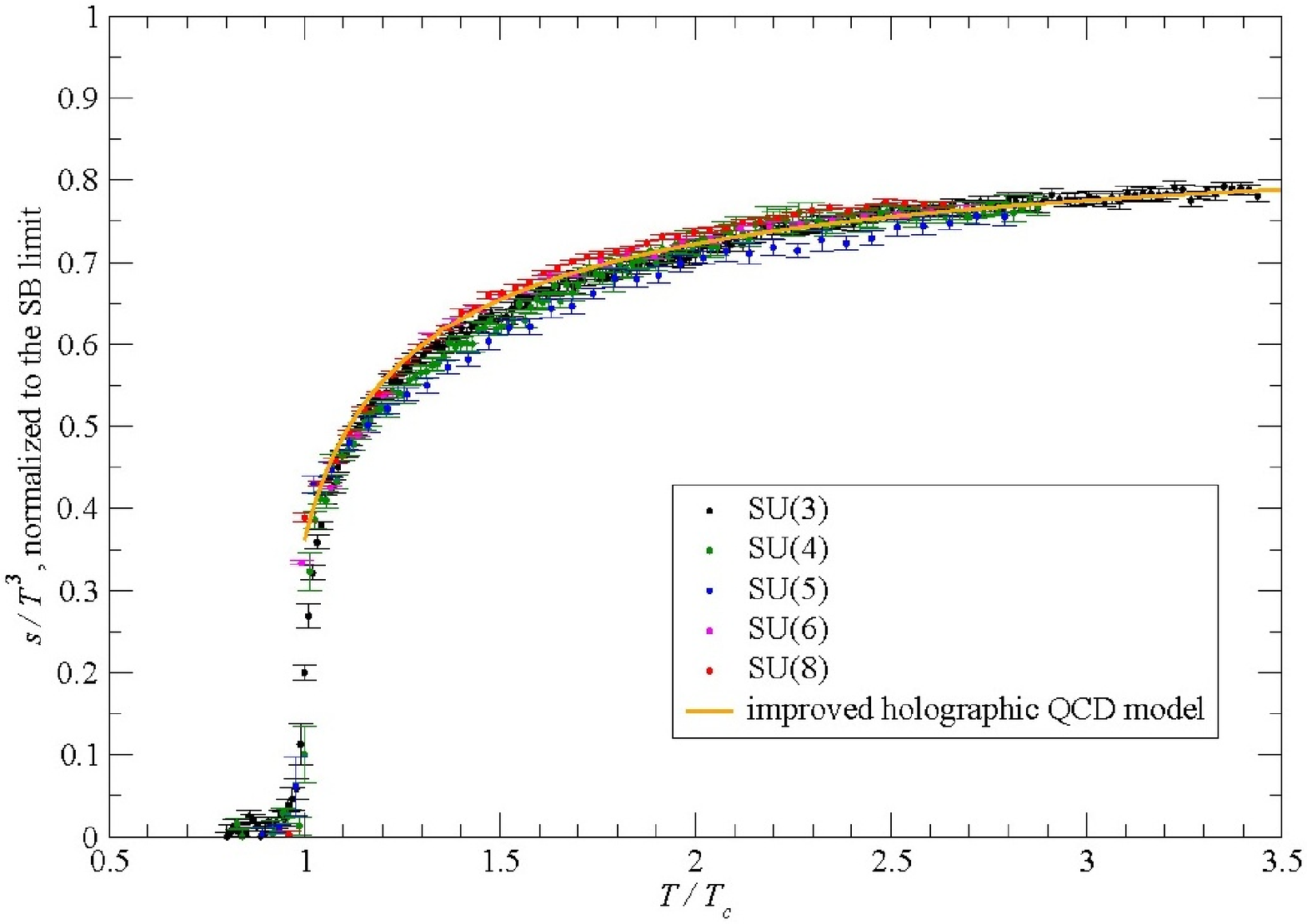} \hspace{-0.7cm}
\includegraphics[width=8.cm]{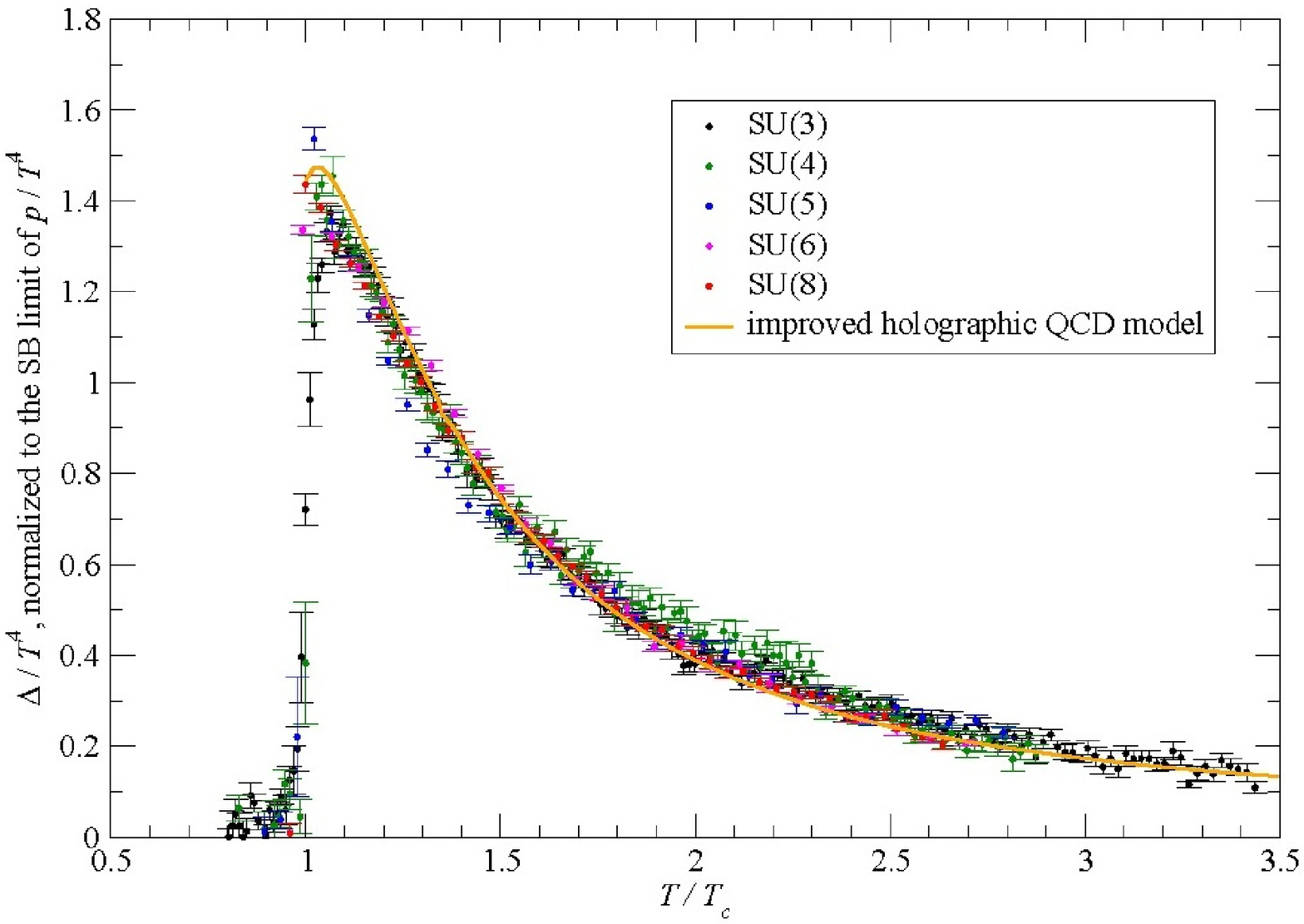}\\
\caption{Left: normalized entropy as function of temperature for YM with various $N_c=3,4,5,6,8$.  Right: the normalized trace of the stress tensor (interaction measure).  The yellow lines are based on  the calculations of IHQCD. Data by M. Panero from \cite{Panero}.}
\label{fig1}
\end{figure}

\begin{figure}[h!]
 \begin{center}
\includegraphics[scale=0.7]{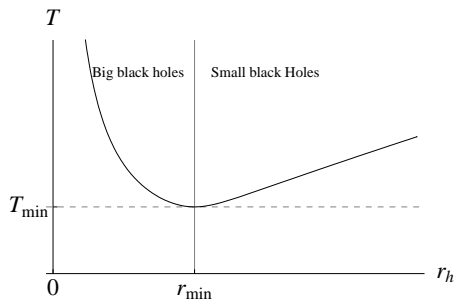}
 \end{center}
 \caption[]{Typical plot of the black-hole temperature
 as a function of the  horizon position $r_h$, in a confining background.
The temperature features a minimum $T_{min}$ at $r_{min}$ ,
that separates the large black-hole  from the small black-hole  branches.
The boundary is at $r=0$ and the IR endpoint at $r=\infty$}
\label{TFrh-BS}
\end{figure}

The theory, upon tuning two phenomenological parameters describes well both zero-temperature spectra of glueballs as well as the thermodynamics at finite temperature, \cite{data}. In particular at $T=0$ the ground state is described by a unique saddle-point solution. For $0<T<T_{min}$, this is the only saddle point available to the system, upon compactification of the time-circle. We will call this solution the ``thermal vacuum solution" as it describes the confining vacuum of large-$N_c$ YM. For $T>T_{min}$, there are generically two saddle point solutions, (beyond the thermal vacuum solution), that are black holes, \cite{gkmn} as can be seen in figure \ref{TFrh-BS}.
The one with the larger horizon, is a black hole with positive specific heat while the smaller one has a negative specific heat. These two black holes merge at $T=T_{min}$.
At $T=T_c>T_{min}$ there is a first order phase transition to the deconfinement phase, described by the large black hole, \cite{gkmn}. At $T>T_{c}$ the theory is in the deconfined phase. A calculation of the thermodynamic functions in this phase, and its comparison to a recent high statistic lattice calculation is shown in figure \ref{fig1}.

The basic characteristics of the translationary invariant black hole solutions of IHQCD are as follows

\begin{itemize}

\item The large black holes have a E(T), S(T) that matches the one in large $N_c$ YM. In particular $E\sim T^4$, $S\sim T^3$ as $T\to \infty$. The horizon position is $r^{\rm large}_h\sim {1\over  T}\to 0 $ at large temperatures, where $r=0$ is the position of the AdS boundary. The specific heat is positive.

\item The small black holes are unstable, with negative specific heat. They are however nowhere near Schwartzschild black holes. Although like Schwartzschild black holes they are unstable, they have scalar hair that makes them have different properties as obvious from their thermodynamic functions.
     As $T\to \infty$ their horizon shrinks to zero size as $r^{\rm small}_h\sim {T\over \Lambda_{QCD}^2}$. Note that $r\to \infty$ is the IR region of the bulk geometry.
    We also have
    \be
    S\simeq V_3~\exp\left[-3{T^2\over  \Lambda_{QCD}^2}\right]\sp E\simeq V_3M_P^3~ T ~\exp\left[-3{T^2\over  \Lambda_{QCD}^2}\right]\sp T\to \infty
    \ee

\end{itemize}

It is interesting that at large $T$, the small and large black holes satisfy the duality relation
\be
(\Lambda_{QCD}~r^{\rm small}_h)(\Lambda_{QCD} ~r^{\rm large}_h)\simeq 1
\ee
which should be compared with the analogous relation in global AdS, $(\ell~r^{\rm small}_h)(\ell ~r^{\rm large}_h)\simeq 1$.

Using as parameter $r_h$, we can put both the large and small branches on the same diagram. For $r_h<r_{\rm min}$ we are in the large bh branch while for $r_h>r_{\rm min}$ we are in the small bh branch. The two branches merge at $r_{min}$ at a common temperature $T_{min}$.
The functions $E(r_h), S(r_h)$ are monotonic (decreasing) functions of $r_h$. As functions of temperature, in each branch, $E(T), S(T)$ are monotonically increasing functions. They also satisfy, \cite{gkmn}
\be
E_{\rm large}(T_i) > E_{\rm small}(T_j) \sp  S_{\rm large}(T_i) > S_{\rm small}(T_j)
\ee
for any $T_i,T_j$.

\section{Black holes in heavy-ion collisions}

Although the above properties are for translationary-invariant black holes, we can envisage finite configurations with such profiles. They will not be stable but will have thermodynamic characteristics that are similar.

A particularly important question is what kind of black hole a heavy-ion collision creates. In a sense this is a question for the microcanonical ensemble, as there is a finite available energy $E_i$ to create the black hole.
The questions is, given the energy, what maximizes the entropy, as this will be the most likely outcome of the process.

Noting that the energy density is a monotonic function of $r_h$, there is a limiting energy density, namely $\e_{\rm min}\equiv \e(r_{min})$ that separates large from small black holes. There is also another benchmark, namely $\e_c\equiv \e(r_{c})>\e_{\rm min}$ at the critical radius, where the first order phase transition takes place.\footnote{In large-$N_c$ YM the transition is strongly first order. SU(3) YM still has a (weaker) first order phase transition. Once massive quarks are included the transition becomes a crossover, but for our purposes the crossover is fast enough so that we t treat it a first order phase transition at the level of accuracy we are aiming at.}

 Therefore, if the energy density  $\e_i$ of the heavy-ion collision is bigger than $\e_c$ we expect the formation of a large finite size bh. If $\e_i<\e_{\rm min}$ it is plausible that a small black hole is formed, although this is issue is rather complicated even in the case of AdS, \cite{aharony,minwalla}.
The case $\e_c>\e_i>\e_{\rm min}$ is murkier as in this case there are two possibilities. Either the creation of a large black hole in a subdominant saddle point, or the evolution of energy without thermalization, in the thermal vacuum solution. At $N_c\to\infty$ the difference of the saddle point energies scales as $N_c^2$ and therefore no black hole is expected to form. However for SU(3) YM we might expect a potentially significant chance of thermalization. In view of the crossover behavior of real QCD, this regime is murky.
However at RHIC and LHC, $\e_i\gg \e_c$.

Given an initial energy density $\e_i>\e_c$, distributed in three volume $V_3$ we expect that very fast it will form a trapped surface, \cite{minwalla2,heller}, a signal of thermalization and of developing a horizon. For $V_3>> \Lambda_{QCD}^{-3}$ as expected in heavy-ion collisions,  a large black hole is
formed with finite spatial extend roughly given by $V_3$. Outside this extend,  the saddle point solution is that of the vacuum solution.

Because of the finite spatial extend this solution is unstable and begins to evolve with time. The main form of evolution is that it ``falls" in the bulk geometry. On one hand there is a "surface tension at the spatial boundaries where there are big gradients, and the black hole expands in them. At the same time the radial horizon position recedes to the IR as expected from a ``falling" energy distribution in the bulk. This keeps that horizon area approximately constant, and is therefore an isentropic evolution.
The way this happens is that the 4d-profile remains invariant in comoving coordinates. It seems shrinking because it moves to larger values of r, where the metric is different.
Therefore, because of the warping, the size in transverse space expands in physical units while the size in the radial direction contracts (in physical
units). This is an adiabatic expansion in 4d QCD terms.

In realistic heavy-ion collisions,
there is deposition of about 20 TeV of energy (RHIC) in the
midrapidity range during the first 0.1-0.5 fm/c with initial   density  about 15 GeV/fm$^3$.
From 0.5 - 1 fm/c there is ultrafast thermalization and entropy production.
The free-fall phase is from about 1-6 fm/c during which the entropy is constant.  Finally in the last 6-10 fm/c this is an isothermal phase dominated by the transition.

During the free-fall phase the Hawking temperature of the black hole is a decreasing function of time. It starts from $\e_i=\e(T_i)$ and it changes while keeping the entropy constant. As the entropy of the large black holes is that of YM, the gravitational and plasma pictures are identical.

We may use the first law to calculate
$ \e'(T)=Ts'(T)$ and  $s'-{3\over T}s={A'\over T}$ with
$A$ the conformal anomaly, $A(T)=\e(T)-3P$.
This relation can be integrated to
\be
s(T)=s_c{T^3\over T_c^3}+T^3\int^{T}_{T_c}A'(u){du\over u}\sp \e(T)={3s_c\over 4}{T^4\over T_c^3}+{3T^4\over 4}\int^{T}_{T_c}A'(u){du\over u}+{A(T)\over 4}
\ee
where $T_c$ is the temperature of the 1rst order phase transition and $S_c$ is the transition entropy in the deconfined phase.
We also obtain $\e_c=\e(T_c)={3\over 4}T_cs_s+{A_c\over 4}$
Below the transition we can approximate $\e\simeq s\simeq A\simeq 0$. Continuity of the free energy implies
$A_c=T_c s_c$,  $\e_c=A_c$.
In Pure YM,
\be
T_c\simeq 240~~{\rm  MeV}\sp  e_c=0.31~N_c^2~T_c^4\simeq (310)^4 ~~{\rm MeV^4}\sp s_c\simeq {\e_c\over T_c}=(338)^3  ~~{\rm MeV^3}
\ee
At the begining of thermalization the energy density is $\e_i\simeq 5.4$ GeV/$fm^3$. (1/fm=200 MeV). The system reaches the cross-over temperature $T_c$ after about $t_{ad}\simeq 6$ fm/c.

For an adiabatic evolution $dS=0$. if we define the linear dimension $L$ as $V_3=L^3$, and as the system expands with the speed of light we have ${dV\over V}={3cdt\over L_0+ct}$. From figure \ref{fig1} we find that ${s(2T_c)\over s(T_c)}\simeq 14$ from where we find that ${V_c\over V_0}=14$ or ${L_c\over L_0}\simeq 2.4$.
From this we find that $ct_{ad}\simeq 1.4 L_0$, or $L_0\simeq 4.3$ fm.

When the black-hole temperature reaches $T_c$, the solution becomes unstable. Unlike the standard Hawking picture of evaporation, the decay here proceeds by bubble nucleation where the confining vacuum is spontaneously appearing in several places seeded by inhomogeneities in the distributions. This process is nearly isothermal, and can be in principle described classically upon bubble nucleation. The decay time can be in principle estimated from the thermodynamic functions of the black hole.

Small black holes on the other hand Hawking evaporate, in the standard fashion, but this process is not relevant for heavy-ion collisions.

It is important to follow this black-hole evolution by solving the Einstein-dilaton equations directly along the lines of \cite{1,2,3,heller}.

\section{Entropy production and particle multiplicities}

The collision of two energy distributions, as realized in heavy-ion collisions, can be approximated at RHIC and LHC energies as the gravitational collision of two shock waves. Following Penrose and Eradley and Giddings we will search for a trapped surface that forms in the part of the space-time that is described by the superposition of two shock waves, before they start interacting. This trapped surface will eventually evolve to become a horizon. The area of this trapped surface therefore will give us  a lower bound on the generated entropy during the collision, which will be converted to a multiplicity for the hadronic final state.  Such an approach was developed recently for the AdS geometry in \cite{11}-\cite{Taliotis}.

The metric and dilaton for such shocks can be found by solving the equations stemming from (\ref{kira1}) with shock wave ansatz
\be
ds^2=b(r)^2\left[{dr^2}+dx^idx^i-2dx^+dx^-+\Phi(r,x^1,x^2)\delta(x^+)(dx^+)^2\right], \qquad \phi = \phi(r,x^+)
 \ee
Compatibility of these equations implies that $\partial_{+}\phi=0$.
The equations for $\Phi$ are

\be
\left(\nabla_{\perp}^2+3\frac{b'}{b}\partial_r+\partial_r^2\right)\Phi =-2\kappa_5^2J_{++}, \hspace{0.15in}\nabla_{\perp}^2\equiv \partial_i\partial_i\sp \kappa_5^2 \equiv 8 \pi G_5
\ee
where we have introduced a stress-tensor source $J_{++}$.

The trapped surface and its boundary $C$, is determined by a function $\Psi$ satisfying the differential equation and boundary conditions
\be
 (\Box_{AdS_3}-A(b(r))) (\Psi-\Phi)=0\sp \Psi|_{C}=0\sp \sum_{i=1,2,r}(\nabla_i\Psi)^2|_C=8b^2
 \ee
We may then compute the area that will provide a lower bound on the entropy
\be
S \geq S_{\rm trapped}=\frac{1}{2G_5} \int_C \sqrt{\det|g_{AdS_3}|}dz~ d^2x_{\perp}=\frac{1}{2G_5} \int_{r_{C_2}}^{r_{C_1}}  b(r)^3 x^2_{\perp}(r)
\ee
where the (generalized) curve $C$ defines the boundary of the trapped surface $S$ and where we have included the two sections of the surface associated with the two shocks.

\subsection{Shocks without transverse dependence}

This is the simplest case of a shock which corresponds to an infinite homogeneous nucleus on the transverse plane. Although the transverse distribution is unrealistic, such a simplification allows for tractable calculations while it describes qualitatively various processes of QCD \cite{Albacete:2008ze,Taliotis:2009ne}. As a first step, we begin our investigation using these kind of shocks where we extract useful hinds on the entropy generation leaving the study of more realistic shock for an upcoming paper.

In this case the shock $\Phi(x^+,r)$ can be determined and the trapped surface ends at $r=r_H$ with $r_H$ determined from
\be
\Phi(r,x^+)=E\delta(x^+)\int{dr\over b^3}
\sp {b^3(r_H)}={E\over \sqrt{8}}
 \label{10}
\ee
with $E\sim s^{1\over 2}$, and $s$ the center of mass energy of the collision.
The area of the trapped surface is
\be
A_{trapped}\simeq \int_{\infty}^{r_H}b^3
\ee
We may therefore estimate the energy dependence of the trapped area for different bulk geometries.
\begin{itemize}

\item For non-confining scaling theories $b\sim r^{-\gamma}$,
with $1\leq \gamma <\infty$. The AdS case corresponds to $\gamma=1$.
We obtain
$A_t\sim s^{3\gamma-1\over 6\gamma}$. This agrees with previous estimates for the AdS case.

\item Confining backgrounds that are scale invariant in the IR, \cite{GK}, with $b(r) \sim (r_0-r)^\delta$, $\delta>{1\over 3}$. In this case we obtain $A_t\sim s^{3\delta+1\over 6\delta}$ at high energy. In this case the exponent varies between ${1\over 2}$ and 1.

 \item Confining backgrounds with $b(r)\sim e^{-(\Lambda r)^a}$. In this case $A_t\sim s^{1\over 2}(\log s)^{a+1\over a}$ at high energy.

      \item Confining backgrounds with $b(r)\sim e^{-\left({\Lambda \over r-r_0}\right)^a}$. In this case $A_t\sim s^{1\over 2}(\log s)^{1-a\over a}$ at high energy.

\end{itemize}

Note that in all cases the entropy production is larger than AdS.

This simple context does not include the realistic structure of the transverse space distributions of real world heavy-ion collisions, but captures basic tenets of the approach. In particular it shows that the assumption of the gravitational description up to the boundary suggests a stronger $s$ dependence of the multiplicity than seen experimentally. This is a consequence of string interactions of gravity up to the boundary.
In a future publication we will analyse realistic collisions with localised energy distributions in the transverse space.

\section{Outlook}

The gravitational description of heavy-ion collisions, is promising to revolutionize the description of both strong coupling physics in QCD and
that of black hole formations and evaporation. We now have concrete and detailed experimental measurements of black-hole formation and evaporation
 that can help analyse in more detail the physics of (QCD) black holes, and give the only tool to calculate thermalization, hydrodynamics and black-hole decay in heavy ion collisions. More analytical and numerical work is needed in order to analyze in detail these effects, but the technical tools seem to currently available.

 Already CMS has seen hints of collective behavior in high-multiplicity pp collisions at LHC. With higher energy, high-multiplicity pp data,  we will have a new arena for5 black-hole formation and evaporation. It is debatable whether the black holes in that case would be of the large kind. If they are "small" they will be even closer (but not identical) to the Schwartzschild -like black holes people expected from higher dimensional theories.

\section{Acknowledgements}

We would like to thank P. Romatschke for participating in early stages of this work, and G. Dvali, Y. Kovchegov for useful conversations.
This work was partially supported by a European Union grant FP7-REGPOT-2008-1-CreteHEP Cosmo-228644 and PERG07-GA-2010-268246.

\end{document}